\documentclass[aps,pra,twocolumn,superscriptaddress]{revtex4-1}

\usepackage{graphicx}
\usepackage{amsmath}
\usepackage{braket}
\usepackage{nicefrac}
\usepackage{units}
\usepackage{ragged2e}
\usepackage{hyperref}
\usepackage{color}
\usepackage{mathrsfs}
\usepackage{tensor}

\begin{document}
	
	\pagecolor{white}

	\title{Anisotropic polarizability of erbium atoms}
	
	\author{J. H. Becher}
	\altaffiliation[Current address:]{Physikalisches Institut, Universit\"at Heidelberg, Im Neuenheimer Feld 226, 69120 Heidelberg, Germany}
	\affiliation{Institut f\"ur Experimentalphysik, Universit\"at Innsbruck, Technikerstra{\ss}e 25, 6020 Innsbruck, Austria}
	\author{S. Baier}
	\affiliation{Institut f\"ur Experimentalphysik, Universit\"at Innsbruck, Technikerstra{\ss}e 25, 6020 Innsbruck, Austria}
	\author{K. Aikawa}
	\altaffiliation[Current address:]{Department of Physics, Graduate School of Science and Engineering, Tokyo Institute of Technology, Meguroku, Tokyo, 152-8550 Japan}
	\affiliation{Institut f\"ur Experimentalphysik, Universit\"at Innsbruck, Technikerstra{\ss}e 25, 6020 Innsbruck, Austria}

	\author{M. Lepers}
	\altaffiliation[Current address:]{Laboratoire Interdisciplinaire Carnot de Bourgogne, CNRS, Universit\'{e} de Bourgogne Franche-Comt\'{e}, Dijon, France}
	\affiliation{Laboratoire Aim\'{e} Cotton, CNRS, Universit\'{e} Paris-Sud, ENS Paris-Saclay,  Universit\'{e}  Paris-Saclay, 91405 Orsay, France}
	\author{J.-F. Wyart}
	\affiliation{Laboratoire Aim\'{e} Cotton, CNRS, Universit\'{e} Paris-Sud, ENS Paris-Saclay,  Universit\'{e}  Paris-Saclay, 91405 Orsay, France}
	\author{O. Dulieu}
	\affiliation{Laboratoire Aim\'{e} Cotton, CNRS, Universit\'{e} Paris-Sud, ENS Paris-Saclay,  Universit\'{e}  Paris-Saclay, 91405 Orsay, France}
	\author{F. Ferlaino}
	\affiliation{Institut f\"ur Experimentalphysik, Universit\"at Innsbruck, Technikerstra{\ss}e 25, 6020 Innsbruck, Austria}
	\affiliation{Institut f\"ur Quantenoptik und Quanteninformation, \"Osterreichische Akademie der
	Wissenschaften, 6020 Innsbruck, Austria}

	\date{\today}
	
	\begin{abstract}
	We report on the determination of the dynamical polarizability of ultracold erbium atoms in the ground and in one excited state at three different wavelengths, which are particularly relevant for optical trapping. Our study combines experimental measurements of the light shift and theoretical calculations. In particular, our experimental approach allows us to isolate the different contributions to the polarizability, namely the isotropic scalar and anisotropic tensor part. For the latter contribution, we observe a clear dependence of the atomic polarizability on the angle between the laser-field-polarization axis and the quantization axis, set by the external magnetic field.  Such an angle-dependence is particularly pronounced in the excited-state polarizability. We compare our experimental findings with the theoretical values, based on semi-empirical electronic-structure calculations and  we observe a very good overall agreement. Our results pave the way to exploit the anisotropy of the tensor polarizability for spin-selective preparation and manipulation.
	\end{abstract}
	
	\maketitle

	\section{INTRODUCTION}

	Ultracold quantum gases provide many different degrees of freedom, which can be controlled to a very high precision. This makes them a reliable and versatile tool to study complex many-body phenomena in the laboratory \cite{RevModPhys.80.885}. Some of those degrees of freedom rely on the interaction between atoms and light. The strength of such an interaction depends on the atomic polarizability, which is a characterizing quantity of the specific atomic species under examination. Over the course of the last decades, tremendous progress has been made to develop theoretical methods and experimental protocols to determine the atomic polarizabilities, $\alpha_{\rm tot}$, with an increasing level accuracy \cite{bonin1997electric,0953-4075-43-20-202001}. With the gained control over quantum systems, the precise determination of $\alpha_{\rm tot}$  became even more fundamental with implications for quantum information processing, precision measurements, collisional physics, and atom-trapping and optical cooling applications. Calculations of $\alpha_{\rm tot}$ require a fine knowledge on the energy-level structure and  transition matrix elements, which is  increasingly complex to acquire with increasing number of unpaired electrons in the atomic species. 
	For instance, alkali atoms with their single valence electron allow a determination of the static atomic polarizability with an accuracy below  $\unit[1]{\%}$ \cite{schwerdtfeger2015table,PhysRevA.92.052513} when the full atomic spectrum is accounted. \\ 
    In the case of the multi-electron lanthanide atoms (Ln), which have been recently brought to quantum degeneracy (ytterbium (Yb) \cite{takasu2005bose,fukuhara2007degenerate}, dysprosium (Dy) \cite{lu2011strongly,lu2012quantum}, erbium (Er) \cite{aikawa2012bose,aikawa2014reaching}), the atomic spectrum can be very dense with a rich zoology of optical transitions from being ultra narrow to extremely broad. Beside Yb with its filled shell, the other Ln show an electron vacancy in an inner and highly anisotropic electronic shell ($4f$ for all Ln beside lanthanum and lutetium), surrounded by a completely filled isotropic $s$ shell. Because of this peculiar electronic configuration, such atomic species are often referred to as {\em submerged-shell atoms} \cite{doi:10.1021/jp0488416,PhysRevA.81.010702}.\\
    Capturing the complexity of Ln challenges spectroscopic approaches and allows for stringent tests of ab-initio calculations \cite{Chu,0953-4075-50-1-014005, PhysRevA.83.032502,PhysRevA.95.062508,lepers2014anisotropic}.   
    Beside being benchmark systems for theoretical models, Ln exhibit special optical properties, opening novel possibilities for control, manipulation, and detection of Ln-based quantum gases \cite{PhysRevA.91.063414, burdick2016long}. One peculiar aspect of magnetic Ln is their sizable anisotropic contribution to the total atomic polarizability, originating from the unfilled $4f$ shell. Particularly relevant is the anisotropy arising from the tensor polarizability. This term gives rise to a light shift, which is quadratic in the angular-momentum projection quantum number, $m_J$, and provides an additional tool for optical spin manipulation, as recently studied in ultracold Dy experiments \cite{Kao:17}. The anisotropy in the polarizability has been observed not only in atoms with large orbital-momentum quantum number but also in large-spin atomic system, such as cromium (Cr), \cite{PhysRevLett.111.185305,Chicireanu2007} and molecular systems \cite{neyenhuis2012anisotropic, PhysRevLett.113.233004, Cornish, reviewOlivier}.  \\ 
    This paper reports on the measurement of the dynamical polarizability in ultracold Er atoms in both the ground state and one excited state for trapping-relevant wavelengths. Our approach allows us to isolate the spherically-symmetric (scalar) and the anisotropic (tensor) contribution to the total polarizability. We observe that the latter contribution, although small in the ground state, can be very large for the excited state. Our results are in very good agreement with electronic-structure calculations of the atomic polarizability, showing a gained control of the atom-light interaction in Er and its spectral properties.

	\section{THEORY OF DYNAMICAL POLARIZABILITY}
	
\begin{figure*}
	\centering
	\includegraphics[width=\textwidth]{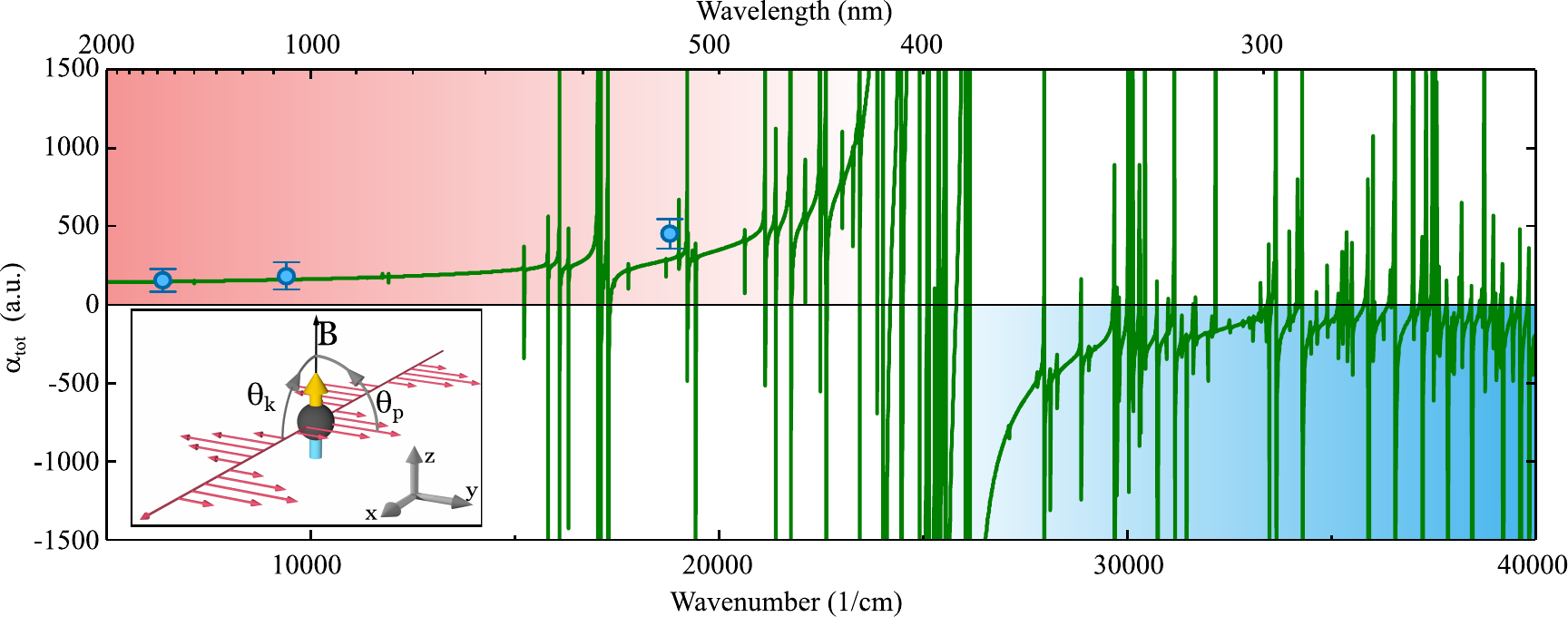}
	\caption{(Color online) Calculated (solid line) and measured (filled circles) atomic polarizability $\alpha_{\rm tot}$ of Er in the ground state for $\theta_p = \theta_k = \unit[90]{^\circ}$ as a function of the light-field wavenumber and wavelength in atomic units. A divergence of the polarizability indicates an optical dipole transition. The finite amplitude of the peaks of the narrow transitions are an artefact caused by the finite number of calculated data points. The red and blue shadows indicate, that there is a broad red-detuned region for long wavelengths without many resonances and also a mostly blue-detuned region in the ultraviolet range. The inset illustrates the configuration of angles $\theta_k$ and $\theta_p$ for the shown data. B denotes the orientation of the magnetic field.\label{Spectrum}}
\end{figure*}

	To understand the concept of anisotropic polarizability, we first review the basic concepts of atom-light interaction \cite{0953-4075-43-20-202001, pub1262}.
When an isotropic medium is submitted to an external electric field, e.g. a linearly-polarized light field, it experiences a polarization parallel to the applied electric field. However, in anisotropic media an external electric field can also induce a perpendicular polarization, which in the atom-light-interaction language corresponds to a polarizability with a tensorial character. As we will discuss in the following, Er atoms can be viewed as an anisotropic medium because of their orbital anisotropy in the ground and excited states (non-zero orbital-momentum quantum number $L\neq 0$). The atomic polarizability is then described by a $3 \times 3$ tensor, $\mathscr{P}$. The total light shift experienced by an atomic medium exposed to an electric field $\vec{E}$ reads as 
\begin{equation}
U = \frac{1}{2}\vec{E}^\dagger  \mathscr{P} \vec{E}.
\label{polatensor}
\end{equation}
Equation.\,(\ref{polatensor}) can be decomposed into three parts. For this we define the scalar polarizability tensor $\mathscr{A}_s$ (diagonal elements), the vectorial polarizability tensor $\mathscr{A}_v$ (anti-symmetric part of the off-diagonal elements) and the tensorial polarizability tensor $\mathscr{A}_t$ (symmetric part of the off-diagonal elements). Hence, a medium with polarizability tensor $\mathscr{P}$ placed into an electric field $\vec{E}$ feels the total light shift
\begin{equation}
U =  \frac{1}{2} \vec{E}^\dagger[\mathscr{A}_s + \mathscr{A}_v + \mathscr{A}_t]\vec{E}.
\label{equtensor}
\end{equation}	
We now consider the case of an atom in its electronic ground state with non-zero angular-momentum quantum number $J$, its projection on the quantization axis $m_J$, and a total polarizability $\alpha_{\rm tot}$ placed in a laser field of intensity $I = \frac{\epsilon_0 c}{2} |\vec{E}|^2$, polarization vector \textbf{u}, and frequency $\omega = 2 \pi \frac{c}{\lambda}$. Here, $\epsilon_0$ is the vacuum permittivity, $c$ is the speed of light and $\lambda$ is the wavelength of the laser field. For a given quantization axis, which is typically set by an external magnetic field, we furthermore define $\theta_k$ $(\theta_p)$ as the angle between the propagation \footnote{In fact, $\theta_k$ is the angle between the vector defined by $\textbf{u}^*\times \textbf{u}$ and the magnetic field. It changes sign when going from right to left circular polarization.} (polarization) axis of the laser field and the quantization axis (see inset in Fig.\,\ref{Spectrum}). As shown in Ref.\,\cite{PhysRevA.95.062508}, the tensor product of Eq.\,(\ref{equtensor}) can be developed and the total light shift can be expressed as the sum of the scalar ($U_s$), vector ($U_v$), and tensor ($U_t$) light shift as follows
\begin{eqnarray}
U(\omega) &=& -\frac{1}{2\epsilon_0 c} I(r) \alpha_{\rm tot}=U_s+U_v+U_t \nonumber\\
&=& -\frac{1}{2\epsilon_0 c} I(r)\Big[\alpha_s(\omega) + |\textbf{u}^*\times \textbf{u}| \cos{\theta_k}\frac{m_J}{2J} \alpha_v(\omega)\nonumber \\
&&	+ \frac{3m_J^2 - J(J+1)}{J(2J-1)} \times \frac{3\cos^2{\theta_p}-1}{2}\alpha_t(\omega)\Big].
\label{potential}
\end{eqnarray}

For convenience, we have explicitly separated the tensor and vector term in two parts. The first part depends on the angles, $J$ and $m_J$, and the second part on $\omega$ and $J$. We refer to the latter as the polarizability coefficients $\{\alpha_s, \alpha_v, \alpha_t\}$ for the scalar, vector, and tensor part, respectively. \\  
Because of their $J$, $\textbf{u}$, and angle dependence, $U_v$ and $U_t$ vanish for special configurations. In particular, $U_v$ vanishes for any linear polarization, since $\textbf{u}^*\equiv\textbf{u}$ is a real vector and thus $|\textbf{u}^*\times \textbf{u}| = 0$ and for elliptical polarization at $\theta_k = \unit[\pm 90]{^\circ}$. $U_t$ vanishes for $\cos{\theta_{p0}} = \sqrt{\nicefrac{1}{3}}$, i.e. for $\theta_{p0} = \unit[54.7]{^\circ}$, or for $J=\nicefrac{1}{2}$. The latter condition is always fulfilled by alkali atomic species, which indeed have zero tensor light shift in the ground state.
As we will discuss later, this is an important difference between alkali and magnetic Ln, such as Dy and Er, which have $J=8$ and $J=6$ in the ground state, respectively. Finally, we note that $U_t$ shows a quadratic dependence on $m_J$, which paves the way for a selective manipulation of individual Zeeman substates.\\
The polarizability coefficients read as  
\begin{eqnarray}
\alpha_s(\omega) &=& -\frac{1}{\sqrt{3(2J+1)}} \alpha_J^{(0)}(\omega) \nonumber \\
\alpha_v(\omega) &=& \sqrt{\frac{2J}{(J+1)(2J+1)}}\ \alpha_J^{(1)}(\omega) \nonumber \\
\alpha_t(\omega) &=& \sqrt{\frac{2J(2J-1)}{3(J+1)(2J+1)(2J+3)}}\ \alpha_J^{(2)}(\omega),
\label{polas}
\end{eqnarray}	
where  $\alpha_J^{(K)}(\omega)$, $K \in \{0,1,2\}$, is known as the coupled polarizability. To precisely calculate the value of the polarizability, it is necessary to know the parameters of each dipole-allowed transition, i.e. the energy of the transition $\hbar \omega_{JJ'}$ and the natural linewidth of the excited state $\gamma_{J'}$. In constant-sign convention \cite{reviewOlivier}, $\alpha_J^{(K)}(\omega)$ is indeed given by a sum-over-state formula over all dipole-allowed transitions ($\Delta J = 0,\pm1$),
\begin{eqnarray}
\alpha_J^{(K)}(\omega) &= \sqrt{2K+1} \times \sum_{J'} (-1)^{J+J'}\nonumber \\
&\begin{Bmatrix} 1 & K & 1 \\ J & J' & J \end{Bmatrix} \lvert\bra{J'}\rvert\textbf{d}\lvert\ket{J}\rvert^2  \times \nonumber \\
& \frac{1}{\hbar} \Re\left[\frac{1}{\Delta_{J'J}^- - i\gamma_{J'}/2} 
+ \frac{(-1)^K}{\Delta_{J'J}^+ - i\gamma_{J'}/2} \right].
\label{sumoverstate}
\end{eqnarray}
Here, $\lvert\bra{J'}\rvert\textbf{d}\lvert\ket{J}\rvert$ is the reduced dipole transition element and $\Delta_{J'J}^\pm = \omega_{J'J} \pm \omega$. The curly brackets denote the Wigner 6-j symbol. Note that the imaginary part of the term in the squared brackets is connected to the off-resonant photon scattering rate. As will be discussed in the next section, a precise knowledge of the atomic spectrum is highly non-trivial for multi-electron atomic species with submerged-shell structure and requires advanced spectroscopic calculations.\\

	\section{ATOMIC SPECTRUM OF ERBIUM}
\begin{figure}
	\centering
	\includegraphics[width=\linewidth]{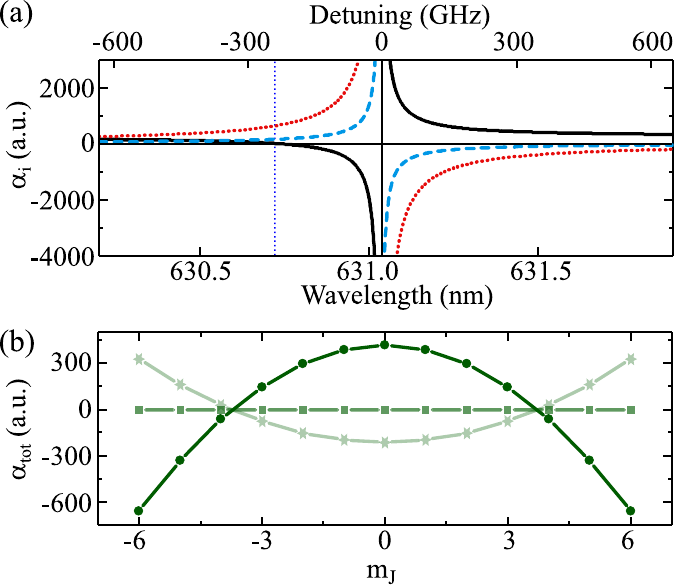}
	\caption{(Color online) Ground-state polarizability of Er in the proximity of a narrow optical transition at $\unit[631.04]{nm}$ with a linewidth of $\unit[2\pi \times 28]{kHz}$. (a) polarizability coefficients $\alpha_s$ (solid line), $\alpha_v$ (dotted line), and $\alpha_t$ (dashed line) versus the laser-field wavelength. The vertical dotted line indicates the zero-crossing of $\alpha_s$. (b) total polarizability $\alpha_{\rm tot}$ as a function of $m_J$, identifying the different Zeeman sub-levels of the ground-state manifold for $\theta_p = \unit[90]{^\circ}$ (circles), $\theta_{p0} = \unit[54.7]{^\circ}$ (squares) and $\theta_p = \unit[0]{^\circ}$ (stars) calculated with Eq.\,(\ref{potential}) for $\theta_k = \unit[90]{^\circ}$ at $\unit[630.7]{nm}$, corresponding to the wavelength of the zero-crossing of $\alpha_s$.}
	\label{631polas}
\end{figure}

	The submerged-shell electronic configurations of Er in its ground state reads as $[\mathrm{Xe}]4f^{12}6s^2$, accounting for a xenon core, an open inner $f$ shell with a two-electron vacancy, and a closed $s$ shell. The corresponding total angular momentum is $J=6$, given by the sum of the orbital ($L=5$) and the spin ($S=1$) quantum number.\\
The calculated static polarizability of ground-state Er is $\unit[149]{a.u.}$ \footnote{This value is obtained using the improved calculations developed in this work and is slightly higher than the value $\unit[141]{a.u.}$ reported in Ref.\,\cite{lepers2014anisotropic}}. To calculate the dynamical one, $\alpha_{\rm tot}(\omega)$, we use  Eq.\,(\ref{potential}) and Eq.\,(\ref{sumoverstate}), based on the semi-empirical electronic-structure calculation from Ref.\,\cite{lepers2014anisotropic}. The result is shown in Fig.\,\ref{Spectrum} for the case of light propagating along the $x$-axis and linearly polarized along the $y$-axis ($\theta_k=\theta_p=\unit[90]{^\circ}$, see Fig.\,\ref{Spectrum} (inset)). Note that for this configuration the vectorial contribution vanishes and the tensor part is maximally negative. The ground-state polarizability of Er is  mainly determined by the strong optical transitions around $\unit[400]{nm}$. The broadest transition is located at $\unit[401]{nm}$ with a natural width of $\unit[2\pi \times 29.7]{MHz}$ \cite{FrischPhD}. Apart from the broad transitions, Er also features a number of narrow transitions. As indicated in the figure by the red-shaded region to the left of the strong resonances, i.e. for wavelengths above $\unit[500]{nm}$, there is a large red-detuned region. To the right, i.e. for wavelengths below $\unit[380]{nm}$, the atomic polarizability is mainly negative (blue-shaded region), which enables the realization of blue-detuned dipole traps for e.g. box-like potentials \cite{PhysRevLett.118.123401}.\\
As shown with Dy \cite{Kao:17}, narrow lines give prospects for state-dependent manipulation of atomic samples. We find that a promising candidate for spin manipulation is the transition coupling the ground state to the $J' = 7$ excited state at $\unit[631.04]{nm}$ with a natural linewidth of $\unit[2\pi\times 28]{kHz}$ \cite{ban2005laser}, which we here investigate theoretically. It is weak enough to allow near-resonant operation with comparatively low scattering rate and features large vector and tensor polarizabilities. Figure\,\ref{631polas}(a) shows the calculated values of $\alpha_s$, $\alpha_v$, and $\alpha_t$ of the ground state in the proximity of this optical transition, calculated with Eq.\,(\ref{polas}) and (\ref{sumoverstate}). Interestingly, $\alpha_s$ has a sign opposite to $\alpha_v$ and $\alpha_t$ and crosses zero around $\unit[630.7]{nm}$, where still very large vector ($\unit[680]{a.u.}$) and tensor ($\unit[175]{a.u.}$) polarizabilities persist. Such wavelengths are very interesting since they allow to freely tune the total light shift by changing the polarization of the laser light. The lower panel in Fig.\,\ref{631polas} shows the total polarizability $\alpha_{\rm tot}$ as a function of $m_J$ calculated with Eq.\,(\ref{potential}) for the three angles $\theta_p \in \{\unit[0]{^\circ},\unit[54.7]{^\circ},\unit[90]{^\circ}\}$ at the zero-crossing of the scalar polarizability for $\theta_k= \unit[90]{^\circ}$. $\alpha_{\rm tot}$ depends quadratically on $m_J$ and can be tuned from positive to negative by changing $\theta_p$ while keeping $\theta_k$ constant. By changing $\theta_k$, the vertex of the parabola in Fig.\,\ref{631polas} can be shifted towards higher or lower values of $m_J$, such that $\alpha_{\rm tot}$ vanishes for a particular $m_J$ state. Such a feature can in principle be used for a state-dependent manipulation or trapping of the atomic sample \cite{yang2017spin}.

	\section{MEASUREMENTS}
To extract the polarizability of Er, we measure the light shift at three wavelengths $\unit[532.26]{nm}$, $\unit[1064.5]{nm}$ and $\unit[1570.0]{nm}$. In addition, we study the polarizability of one excited state, located at $\unit[17157]{cm^{-1}} \equiv \unit[583]{nm}$ with respect to the ground state for $\unit[1064.5]{nm}$ and $\unit[1570.0]{nm}$. This optical line is particularly relevant for ultracold Er experiments, since it is used as the laser cooling transition in magneto-optical traps (MOT).\\
For the measurements, we initially cool down a sample of $^{168}$Er in a MOT \cite{frisch2012narrow}. Here, the atoms are spin polarized to the lowest level of the ground-state Zeeman manifold ($J=6$, $m_J = -6$). We then transfer the sample into a crossed-beam optical dipole trap at $\unit[1064]{nm}$. We force evaporation by decreasing the power of the trapping laser following the procedure reported in \cite{aikawa2012bose} and cool the sample down to temperatures of several $\mu K$.

\subsection{Measurement of the ground-state polarizability}

For the measurement of the polarizability at $\omega=2\pi c/\lambda$, we load the thermal sample  from the crossed-beam dipole trap into an optical dipole trap generated by a single focused beam, operating at the desired wavelength $\lambda$. Typical beam waists range from $\unit[18]{\mu m}$ to $\unit[46]{\mu m}$. In this single-beam trap, the thermal sample reaches typical peak densities ranging from $\unit[10^{13}]{cm^{-3}}$ to $\unit[10^{14}]{cm^{-3}}$ and temperatures of several $\unit[]{\mu K}$. The propagation direction of the beam is illustrated in the inset of Fig.\,\ref{Spectrum}, i.\,e.\,with a magnetic field oriented along the $z$-axis and $\theta_k = \theta_p = \unit[90]{^\circ}$.

We extract the corresponding light shift of the ground state by employing the standard technique of trap-frequency measurements. From the trapping frequencies, we infer the depth of the optical potential $U$, which in turn is related to $\alpha_{\rm tot}$ by Eq.\,(\ref{potential}). In harmonic approximation, for a Gaussian beam of power $P$, which propagates along the $x$-axis with elliptical intensity profile $I(y,z) = I_0 \exp{\left(-\frac{2y^2}{w_{y}} - \frac{2z^2}{w_{z}}\right)}$, beam waists $w_{y}$ and $w_{z}$, and $I_0 = \frac{2P}{\pi w_{y}w_{z}}$, the depth of the induced dipole potential $U_0$ is related to the radial trapping frequencies by $\omega_i = \sqrt{-4U_0/\left(w_{i}^2 m\right)}$, where $i\in\{y,z\}$. $m$ is the atomic mass, and $U_0 = - \frac{1}{2 \epsilon_0 c} \alpha_{\rm tot}(\omega) I_0$. By combining the above expressions, we find the relation
\begin{equation}
\omega_i = \sqrt{\frac{4 \alpha_{\rm tot} P}{\epsilon_0 c \pi w_{y}w_{z}w_{i}^2 m}}.
\label{trapfreq}
\end{equation}
In Eq.\,(\ref{trapfreq}), $\alpha_{\rm tot}$ is the only free parameter since we independently measure the $w_{i}$ and $P$ as discussed later. 

We measure the radial trapping frequencies along the $y$ and the $z$-axis by exciting center-of-mass oscillations and monitoring the time evolution of the position of the atomic cloud in time-of-flight images. To excite the center-of-mass oscillation, we instantly switch off the trapping beam for several hundreds of $\mu s$ \footnote{The exact timing depends on the trap parameters.}. During this time the atoms move due to gravity and residual magnetic field gradients. When the trapping beam is switched on again, the cloud starts to oscillate in the trap and we probe the oscillation frequencies $\nu_{z} = \omega_{z}/2 \pi$ along the $z$-axis and $\nu_{y} = \omega_{y}/2 \pi$ along the $y$-axis. 
In order to extract $\alpha_{\rm tot}$ from Eq.\,(\ref{trapfreq}), we precisely measure the beam waists $w_y$ and $w_z$. The most reliable measurements of the beam waists are performed by using the knife-edge method \footnote{We use a mirror directly in front of the viewport of our science chamber and reflect the trapping beam such that we perform the measurement of the waists as close to the atomic sample as possible. We measure the beam diameter at different positions along the beam with a knife edge method \cite{Khosrofian:83} and extract the waist with a fit to the measured beam diameters.}. We measure the beam waists with an uncertainty of the order of $\unit[1]{\%}$. Aberrations and imperfections of the trapping beams however introduce a systematic uncertainty in the measurement of the beam waists.  We estimate a conservative upper bound for such an effect of $\unit[2]{\mu m}$, which provides the largest source of uncertainty in the measurement of the polarizability. The corresponding systematic errors on $\alpha_{\rm tot}$ is up to about $\unit[35]{\%}$. We measure the trap frequencies as a function of the laser powers $P$ and we fit Eq.\,(\ref{trapfreq}) to the measured frequencies, leaving $\alpha_{\rm tot}$ the only free fitting parameter.   \\	
\begin{table*}
		\begin{ruledtabular}
		\begin{tabular}{l c c c c c c c r}
			E $\unit[]{(cm^{-1})}$& $\lambda $ (nm) & $\alpha_{\rm tot}^{\exp}$ (a.u.) & $\alpha_{\rm tot}^{\rm th}$ (a.u.) & $\alpha_s^{\rm th}$ (a.u.) & $\kappa_0^{\rm exp}$ (\%) & $\kappa_0^{\rm th}$ (\%) & $\alpha_t^{\rm exp}$ (a.u.) & $\alpha_t^{\rm th}$ (a.u.)\\	\hline
			$0$ & $532.26$ & $(430 \pm 8_{\rm st} \pm 80_{\rm sys})$ & $317$ & $308$ & $(-5.3 \pm 1)$ &$-9.2$ & $(-15 \pm 3_{\rm st} \pm 6_{\rm sys})$ & $-19$\\
			$0$ & $1064.5$ & $(166 \pm 3_{\rm st} \pm 61_{\rm sys})$ & $176$ & $173$ & $(-1.8 \pm 0.8)$ & $-4.7$& $(-1.9 \pm 0.8_{\rm st} \pm 1.2_{\rm sys})$ & $-5.4$\\
			$0$ & $1570.0$ & $(163 \pm 9_{\rm st} \pm 36_{\rm sys})$ & $162$ & $159$ & - & $-4.1$ & - & $-4.3$\\ \hline
			 & & $\alpha_{s}^{\rm exp}$ (a.u.)& & & & & & \\ \hline
			$17157$ & $1064.5$ & $(66.6 \pm 0.5_{\rm st}\pm 28_{\rm sys})$ & & $91$ & $(-25.6 \pm 1.6)$ & $-29.7$& $(-11.3 \pm 0.5_{\rm st} \pm 2.0_{\rm sys})$ & $-18$ \\ 
			$17157$ & $1570.0$ & $(-203 \pm 9_{\rm st} \pm 50_{\rm sys})$ & & $-254$ & $(104 \pm 6)$& $40.4$ & $(-141 \pm 9_{\rm st} \pm 19_{\rm sys})$ & $-68.5$ \\
		\end{tabular}
		\end{ruledtabular}
	\caption{Experimental and theoretical polarizabilities for Er of the ground state ($\unit[0]{cm^{-1}}$) and of the $\unit[583]{nm}$-excited state ($\unit[17157]{cm^{-1}}$) for three laser wavelengths $\lambda$. $\alpha_{\rm tot}$ for experiment and theory is given for the case $\theta_p=\theta_k=90^\circ$. The relative change of the light shift $\kappa_0$ (see text) and the tensor polarizability coefficient $\alpha_t$ for the ground state and for the excited state are displayed. The polarizability is given in atomic units. To convert atomic units into SI units, use a factor of $\alpha\unit[]{[Hz/\left(W mm^{-2}\right)]} = \alpha\unit[]{[a.u.]} \times 1.6488\cdot 10^{-35}/2h\epsilon_0 c$. For $\alpha_s^{\rm exp}$ we give statistical and systematic errors respectively (see text).}
	\label{table}	
\end{table*}

We apply the above-described procedure to three different wavelengths of the trapping beam. The experimental and theoretical values for $\alpha_{\rm tot}$ are summarized in Table \ref{table}. For completeness, we also give $\alpha_{s}^{\rm th}$. Comparatively speaking, at a wavelength of $\unit[1064.5]{nm}$ we find that Er, as other Ln, exhibits a weaker polarizability as compared for instance to alkali atoms (e.g. $\unit[687.3(5)]{a.u.}$ (calculated) for rubidium \cite{PhysRevA.86.033416}). This is related to the submerged-shell electronic structure of Er and the so called "lanthanide contraction", resulting into valence electrons being more tightly bound to the atomic core, and so more difficult to polarize, than the single outermost electron of alkali atoms \cite{cotton2013lanthanide,lepers2014anisotropic}.\\
The comparison between the measured and calculated values shows an overall very good agreement, especially at $\lambda=\unit[1064.5]{nm}$ and $\unit[1570]{nm}$. In this wavelength region, there are very sparse and weak optical transitions and the polarizability approaches its static value; see Fig.\,\ref{Spectrum}. At $\lambda=\unit[532.26]{nm}$, we observe a larger deviation between experiment and theory. This can be due to the larger density of optical resonances in this wavelength region. Here, the calculated value of $\alpha_s$  is thus much more sensitive to the precise parameters of the optical line (i.\,e.\,energy position and strength).  In addition, our theoretical model predicts a very narrow transition at $\unit[18774]{cm^{-1}} \equiv \unit[532.7]{nm}$ with a linewidth of $\gamma_{J'} = \unit[6.2 \times 10^3]{s^{-1}}$.\\
We point out that, as a result of our improved methodology to calculate transition probabilities, the theory value of $\alpha_s=\unit[173]{a.u.}$ at $\lambda=\unit[1064.5]{nm}$  is slightly larger than the one previously reported in \cite{lepers2014anisotropic}. In particular, our present calculations use a refined value of the scaling factor on mono-electronic transition dipole moments [Er$^+$] \cite{Proposal}, which is now equal to 0.807.

\begin{figure}
	\centering
	\includegraphics[width=\linewidth]{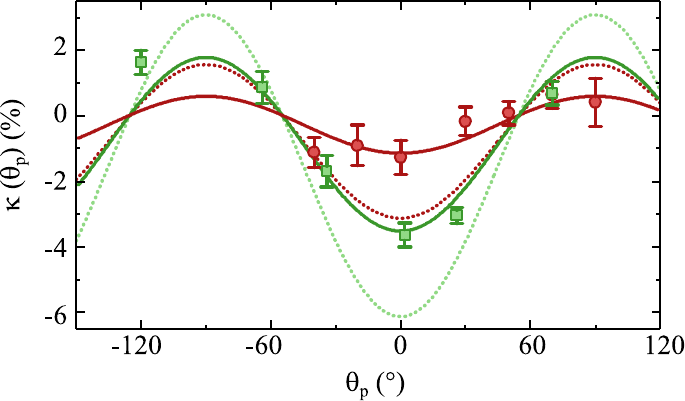}
	\caption{(Color online) Anisotropic polarizability of Er atoms in the ground state. The plot shows the relative change of the light shift at $\unit[532.26]{nm}$ (squares)  and $\unit[1064.5]{nm}$ (circles) for $\theta_k = \unit[90]{^\circ}$ as a function of $\theta_p$. The variation of the total light shift unambiguously reveals the tensor polarizability, which vanishes for an angle of $\theta_{p} \approx \unit[54.7]{^\circ}$. The lines are fits to the data with Eq.\,(\ref{potential}). The error bars indicate the statistical uncertainties from the trapping-frequency measurements. The dotted lines represent the theory prediction.}
	\label{results}
\end{figure}

As previously discussed, Ln exhibit an anisotropic light shift, arising from the sizable tensor contribution to the total polarizability (see Eq.\,(\ref{potential})). This distinctive feature has been experimentally observed in Dy in the proximity of a narrow optical transition \cite{Kao:17}. 
Here, we address this aspect with Er atoms by measuring the light shift in the ground state and its angle dependence at $\unit[532.26]{nm}$ and $\unit[1064.5]{nm}$. At these wavelengths, our theory predicts that $\alpha_t$ for the ground state is of the order of a few percent of $\alpha_s$. To isolate this small contribution and to clear the systematic uncertainties, which could potentially mask the effect, we probe the tensor-to-scalar polarizability ratio as follows. We first prepare the ultracold Er sample in the lowest Zeeman sublevel ($m_j=-6$) in the optical trap, operated at the desired wavelength. We then extract the angle-dependent light shift by repeating the measurements of the trap frequencies for different values of $\theta_p$. This is done by either rotating the magnetic field, while keeping an horizontal polarization of the trapping light, or by rotating the polarization axis of the trapping light at a constant magnetic field. In both measurements we choose $\theta_k = \unit[90]{^\circ}$ such that the vector light shift vanishes. Hence, the total light shift comes only from $\alpha_s$ and $\alpha_t$. Since the scalar light shift is independent of $\theta_p$, a dependence of the total light shift on $\theta_p$ is only caused by $\alpha_t$. We quantify this variation by the relative change of the light shift,
\begin{eqnarray}
\kappa(\theta_p) &=& \frac{U - U_s}{U_s} = \frac{U_t}{U_s} = \frac{\omega(\theta_p)^2 - \omega(\theta_{p0})^2}{\omega(\theta_{p0})^2}\nonumber\\ 
&=&\frac{3m_J^2 - J(J+1)}{J(2J-1)} \times \frac{3\cos^2{\theta_p}-1}{2} \frac{\alpha_t}{\alpha_s}.
\label{kappa}
\end{eqnarray}
Note that the first factor in the second line of Eq.\,(\ref{kappa}) is equal to one for $\ket{J,m_J} = \ket{6,-6}$, such that the peak-to-peak variation of $\kappa(\theta_p)$ corresponds to $\kappa_0 = 1.5 \times \frac{\alpha_t}{\alpha_s}$. Figure\,\ref{results} shows $\kappa(\theta_p)$ for $\unit[532.26]{nm}$ and $\unit[1064.5]{nm}$. At both wavelengths, the data shows the expected sinusoidal dependence of $\kappa$ on $\theta_p$. We fit Eq.\,(\ref{kappa}) to the data and extract $\kappa_0$ and $\alpha_t$. Our results are summarized in Table\,\ref{table}. The systematic uncertainties of $\alpha_t$ are obtained by error propagating the systematical errors of $\alpha_s$. 
We observe that $\alpha_t$ for the ground-state gives only a few percent contribution to the total atomic polarizability. However, the corresponding tensor light shift for the typical power employed in optical trapping can already play an important role in spin-excitation phenomena in Er quantum gases \citep{FlipFlop}. 

Given the complexity of the Er atomic spectrum and the small tensorial contribution, it is remarkable the good agreement between the theoretical predictions of $\alpha_t$ and the experimental value for both investigated wavelengths. The slightly smaller values extracted from the experiments can be due to additional systematic effects in the measurements. For comparison, we note that at $\unit[1064]{nm}$, $\kappa_0$ for ground-state Er is slightly larger than the one for Dy, which was predicted to be around $\kappa_{0,\rm Dy}^{\rm th} = \unit[1.1]{\%}$ \cite{0953-4075-50-1-014005}, and larger than the one of Cr atoms, which was calculated to be $\kappa_{0,\rm Cr}^{\rm th} = \unit[0.5]{\%}$ (at $\unit[1075]{nm}$) \cite{Chicireanu2007} but was then measured to be significantly lower \cite{PhysRevLett.111.185305}. In Cr experiments, the tensorial contribution to the total light shift was then enhanced by using near-resonant light.

\subsection{Measurement of the excited-state polarizability}

\begin{figure}
	\centering
	\includegraphics[width=\linewidth]{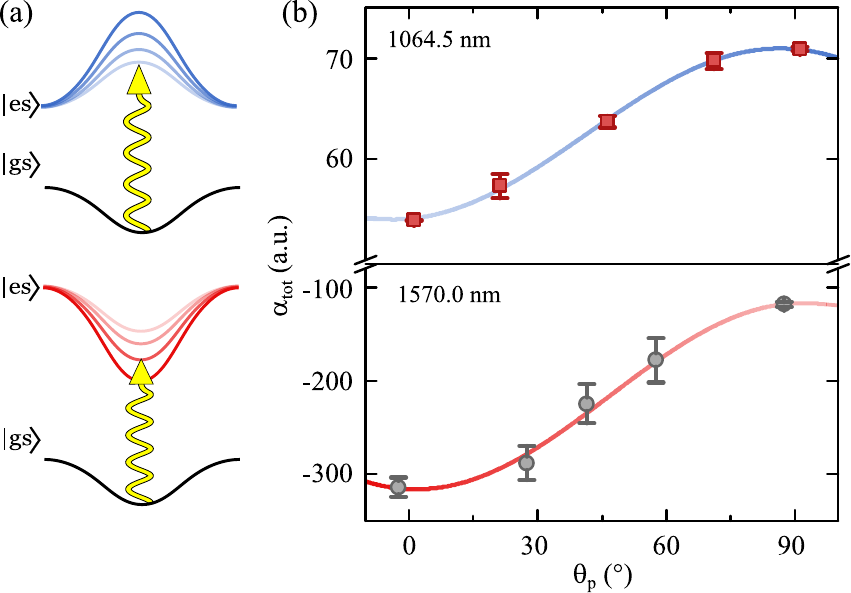}
	\caption{(Color online) $\unit[583]{nm}$-excited-state polarizability. (a) illustration of the energy of atoms in an optical dipole trap with gaussian shape. The upper (lower) panel indicates the case with the excited-state polarizability negative (positive). We measure the shift of the bare atomic resonance in the optical dipole trap (see text) for different values of $\theta_p$ (dark to light red and light to dark blue). This shift is given by the sum of the light shifts in the ground and in the excited state ($\ket{J,m_J} = \ket{6,-6} \rightarrow \ket{J',m_{J'}} = \ket{7,-7}$). To extract the excited-state light shift, we subtract the ground-state shift. (b) $\unit[583]{nm}$-excited-state polarizability for $\unit[1064.5]{nm}$ (red squares) and for $\unit[1570.0]{nm}$ (grey circles). The solid lines indicate fits to the data.}
	\label{583nm}
\end{figure}

 Although small in the ground state, $\alpha_t$ is expected to be substantially larger in the excited state. Therefore, measuring the $\unit[583]{nm}$-excited-state polarizability provides a further test of the level calculations. To extract the excited-state polarizability, we measure the shift of the atomic resonance in the dipole trap. As is depicted in Fig.\,\ref{583nm}(a), the dipole trap induces a light shift not only to the ground state but also to the excited state. To measure the excited-state light shift, we prepare the atomic sample as above described and apply a short pulse of a circularly-polarized probe light at $\unit[583]{nm}$ to the sample. This light couples the ground-state $\ket{J,m_J} =\ket{6,-6}$ level to the $\ket{J',m_J'} =\ket{7,-7}$ sub-level of the excited state manifold of energy $\unit[17151]{cm^{-1}}$  ($[\mathrm{Xe}]4f^{12}6s6p (^3P_1)$). We find a resonant atom loss when the frequency of the probe light matches the energy difference between the ground and the excited state. By scanning the frequency of the probe light, we extract the resonance frequency. This frequency is shifted from that of the bare optical transition by the sum of the ground-state polarizability and the excited-state polarizability. Subtracting the ground-state shift reveals the light shift of the excited state. For this we use the here reported experimental values of the ground-state polarizability and neglect the angle dependence thereof since its anisotropy is two orders of magnitude smaller than the anisotropy of the excited state. We repeat this measurement for various values of $\theta_p$ and find a large angle dependence as we show in Fig.\,\ref{583nm}(b) for $\unit[1064.5]{nm}$ and $\unit[1570]{nm}$. This is expected due to the highly anisotropic wavefunction of the $6p$ electron in the $\unit[583]{nm}$ excited state. From our data, similarly to the ground-state measurements, we extract both the scalar and the tensor polarizability coefficients. The results and the theoretical calculations are presented in the lower section of Table\,\ref{table}. The scalar polarizability coefficient agrees within the error with the theoretical expectations indicating a good understanding of the excited state polarizability. The tensor polarizabilitiy coefficients qualitatively match well with the theoretical values. The quantitative disagreement by up to a factor of two is probably caused by uncertainties in the parameters of strong transitions close by. 

\section{CONCLUSION AND OUTLOOK}
In this paper we presented measurements of the scalar and tensor polarizability of Er atoms in the ground and the $\unit[583]{nm}$-excited state for three wavelengths. Our results qualitatively agree with our theoretical calculations of the polarizability and prove a good understanding of the level structure of Er. A similarly comprehensive picture of the correspondence between theoretical and experimental values of polarizability in Dy is still pending \cite{lu2011strongly,maxenceDy,Kao:17}.\\
For $\unit[1064.5]{nm}$ and $\unit[1570.0]{nm}$ we find excellent agreement of the scalar polarizability. For $\unit[532.26]{nm}$ we observe that the measured value of $\alpha_s$ deviates from the calculated value, which we attribute to the proximity to optical transitions. The measured tensor polarizabilities at $\unit[532.26]{nm}$ and $\unit[1064.5]{nm}$ are of the order of few percent with respect to the scalar polarizabilities and qualitatively agree with the theoretical values.\\
The polarizability of the $\unit[583]{nm}$-excited state was measured to be positive (negative) for $\unit[1064.5]{nm}$ $(\unit[1570]{nm})$, in agreement with the theory. Further it shows a large anisotropy due to the highly anisotropic electronic configuration around the core. Our measured values qualitatively agree with the calculations.\\
As was discussed, the anisotropic polarizability does not only depend on the angle between the quantization axis and the polarization of the light but also gives rise to a $m_J$ dependence of the total light shift. This can be of great importance for experiments with Ln, since it allows for the deterministic preparation or the manipulation of spin states or for the realization of state or species-dependent optical dipole traps.

\begin{acknowledgments}
We thank  R. Grimm, M. Mark, E. Kirilov, G. Natale and V. Kokouline for fruitful discussions.
M. L., J.-F. W. and O. D. acknowledge support from "DIM Nano-K" under the project "InterDy", and from "Agence Nationale de la Recherche" (ANR) under the project "COPOMOL" (Contract No. ANR-13-IS04-0004-01).
The Innsbruck group is supported through an ERC Consolidator Grant (RARE, no.\,681432) and a FET Proactive project (RySQ, no.\,640378) of the EU H2020. The Innsbruck group thanks the DFG/FWF (FOR 2247).\\
\end{acknowledgments}

* Correspondence and requests for materials
should be addressed to F.F.~(email: Francesca.Ferlaino@uibk.ac.at).

** J.H.B. and S.B. contributed equally to this work.

\end{document}